\documentclass[twocolumn]{aastex63}

\newcommand{\swift}{\textit{Swift }}

\usepackage{amsmath}

\submitjournal{AJ}

\shorttitle{A \textit{Swift} Fix}
\shortauthors{Hinkle et al.}

\graphicspath{{./}{figures/}}

\begin{document}

\title{A \textit{Swift} Fix for Nuclear Outbursts}

\correspondingauthor{Jason T. Hinkle}
\email{jhinkle6@hawaii.edu}

\author[0000-0001-9668-2920]{Jason T. Hinkle}
\affiliation{Institute for Astronomy, University of Hawai`i, 2680 Woodlawn Drive, Honolulu, HI 96822, USA}

\author[0000-0001-9206-3460]{Thomas W.-S. Holoien}
\altaffiliation{NHFP Einstein Fellow}
\affiliation{The Observatories of the Carnegie Institution for Science, 813 Santa Barbara Street, Pasadena, CA 91101, USA}

\author[0000-0003-4631-1149]{Benjamin. J. Shappee}
\affiliation{Institute for Astronomy, University of Hawai`i, 2680 Woodlawn Drive, Honolulu, HI 96822, USA}

\author[0000-0002-4449-9152]{Katie~Auchettl}
\affiliation{School of Physics, The University of Melbourne, Parkville, VIC 3010, Australia}
\affiliation{ARC Centre of Excellence for All Sky Astrophysics in 3 Dimensions (ASTRO 3D)}
\affiliation{Department of Astronomy and Astrophysics, University of California, Santa Cruz, CA 95064, USA}

\begin{abstract}
\noindent In November 2020, the \swift team announced an update to the  UltraViolet and Optical Telescope calibration to correct for the loss of sensitivity over time. This correction affects observations in the three near ultraviolet (UV) filters, by up to 0.3 mag in some cases. As UV photometry is critical to characterizing tidal disruption events (TDEs) and other peculiar nuclear outbursts, we re-computed published \swift data for TDEs and other singular nuclear outbursts with \swift photometry in 2015 or later, as a service to the community. Using archival UV, optical, and infrared photometry we ran host SED fits for each host galaxy. From these, we computed synthetic host magnitudes and host-galaxy properties. We calculated host-subtracted magnitudes for each transient and computed blackbody fits. In addition to the nuclear outbursts, we include the ambiguous transient ATLAS18qqn (AT2018cow), which has been classifed as a potential TDE on an intermediate mass black hole. Finally, with updated bolometric light curves, we recover the relationship of \citet{hinkle20a}, where more luminous TDEs decay more slowly than less luminous TDEs, with decreased scatter as compared to the original relationship.
\end{abstract}


\section{Introduction} \label{sec:intro}
A tidal disruption event (TDE) occurs when a star passes too close to a supermassive black hole (SMBH) and is torn apart by tidal forces. A fraction of the disrupted stellar material is subsequently accreted onto the SMBH, resulting in a short-lived, luminous flare \citep[e.g.,][]{lacy82,rees88,evans89,phinney89}. Because they can occur in quiescent galaxies, TDEs are useful as probes of inactive black holes, and allow the study of accretion disks as they form, evolve, and are disrupted on observable timescales.

In recent years, wide-field, untargeted transient surveys, such as the All-Sky Automated Survey for SuperNovae \citep[ASAS-SN;][]{shappee14, kochanek17}, the Asteroid Terrestrial Last-impact Alert System \citep[ATLAS;][]{tonry18}, the Panoramic Survey Telescope and Rapid Response System \citep[Pan-STARRS][]{chambers16}, and the Zwicky Transient Facility \citep[ZTF;][]{bellm19} have put substantial effort into identifying and studying transient outbursts thought to be TDEs. TDEs are identified as discrete flares nuclear flares with hot ($1-5\times10^4$~K) blackbody spectral energy distributions (SEDs) and broad hydrogen and/or helium lines in their optical spectra. Their temporal evolution is very different from the typical stochastic variability of AGN. In the process of searching for TDEs, other nuclear outbursts have also been discovered. Some may be unusual TDEs, such as TDEs occurring around existing AGN \citep[e.g.,][]{blanchard17, payne20}, TDEs caused by an intermediate or stellar-mass black hole \citep[e.g.,][]{perley19, kremer20}, or unrelated phenomena like ``rapid turn-on'' and changing-look AGN \citep[e.g.,][]{shappee14, wyrzykowski17, frederick19, trakhtenbrot19a}. 

A common feature of these nuclear outbursts is that a significant portion of their emission is emitted at ultraviolet (UV) wavelengths. As such, observations from space-based UV telescopes, in particular the \textit{Neil Gehrels Swift Observatory} \citep[\textit{Swift};][]{gehrels04}, are crucial for characterizing the  temperatures and luminosities of these events. Nearly all transients identified as possible TDEs have thus been the subjects of extended monitoring campaigns with the \swift UltraViolet and Optical Telescope \citep[UVOT;][]{roming05}.

In Novemeber 2020, the \swift team announced that due to a loss of sensitivity over time, the photometric calibration for the three UVOT UV filters needed to be retroactively corrected\footnote{\url{https://www.swift.ac.uk/analysis/uvot/index.php}}. This loss of sensitivity can affect UV observations made with \swift by up to 0.3 magnitudes. The most recent correction files indicate that the sensitivity calibration was over-approximated for all three UV filters beginning in late 2015, reaching a $\sim 5$\% level in 2017. Since \swift observations are often used to estimate blackbody temperatures in nuclear outbursts, a difference of 0.3 mag can have a significant effect on the estimated blackbody temperatures and luminosities, particularly for cases where the transient magnitude is close to that of its host galaxy. The UVOT photometry correction thus has the potential to affect not only the conclusions about individual objects, but also the conclusions of population studies \citep[e.g.,][]{arcavi14, hung17, hinkle20a, vanvelzen20}.

Here we re-compute the \swift photometry for all previously published epochs of \swift data taken of TDEs and other singular nuclear outbursts that were observed by \swift in 2015 or later. This includes both transients discovered after 2015 and those discovered prior to 2015 that were still being observed after 2015. We have also used multi-wavelength archival data to model the SEDs of the transient host galaxies and produce host-subtracted light curves and blackbody models of these transients in a uniform way. We present the resulting corrected \swift light curves, host-subtracted light curves, and blackbody models, which we make publicly available.

In Section~\ref{sec:sample} we discuss the sample selection. Section~\ref{sec:host_fits} discusses the sources of archival photometry and the models for the SEDs of the transient host galaxies. Section~\ref{sec:swift_reductions} discusses our reduction of the \swift UVOT data and presents the raw and host-subtracted \swift light curves. Section~\ref{sec:bb_fits} covers our blackbody models of the transient SEDs and presents the resulting luminosities, radii, and temperatures. Section~\ref{sec:L40} discusses our re-analysis of the peak-luminosity/decline-rate relationship we first presented in \citet{hinkle20a}. Finally, Section~\ref{sec:summary} summarizes the results of this work. Throughout this paper, we have used a cosmology with $H_0$ = 69.6 km s$^{-1}$ Mpc$^{-1}$, $\Omega_{M} = 0.29$, and $\Omega_{\Lambda} = 0.71$. 

\section{Sample}
\label{sec:sample}

For our re-analysis, we selected 38 objects that have been classified as TDE candidates or other nuclear outbursts in the literature. These objects are listed in Table~\ref{tab:sample}, along with the references where the \swift photometry was originally published and the source classifications, typically either as an AGN or a TDE. Where the nature of a source is unclear, we list the classification as an ``AGN/TDE''. All but one of our sources is consistent with the host nucleus. The lone exception, ATLAS18qqn (AT2018cow), is the brightest of the growing class of fast optical transients \citep[e.g.,][]{prentice18, ho20, coppejans20}. It has been interpreted as either an exotic supernova \citep[e.g.,][]{prentice18, perley19}, the tidal disruption of a star by an intermediate mass black hole \citep[e.g.,][]{perley19, uno20}, or the tidal disruption of a star by a stellar mass black hole in a star cluster \citep{kremer20}. We list it as ``Ambiguous'' in Table \ref{tab:sample} and include it in our sample due to the potential TDE classification.

In our \swift re-analysis, we used a 5\farcs{0} radius aperture except for ASASSN-17cv, ASASSN-19dj, ZTF18aajupnt, ZTF19aaiqmgl, and ZTF19abzrhgq which used 10\farcs{0}, 15\farcs{0},  15\farcs{0}, 10\farcs{0}, and 10\farcs{0} radius apertures respectively. These larger apertures were chosen to incorporate the entire host galaxy. Additionally, the transient photometry for ATLAS18qqn was measured using a 3\farcs{0} radius aperture to minimize host contamination as the source is non-nuclear. We chose 5\farcs{0} as the default for sources because 5\farcs{0} is the standard \swift aperture radius and has small aperture loss corrections \citep{poole08}.

\begin{deluxetable*}{cccccc}[htb!]
\tabletypesize{\footnotesize}
\tablecaption{Sample of Objects}
\tablehead{
\colhead{Object} &
\colhead{TNS ID} &
\colhead{Right Ascension} &
\colhead{Declination} &
\colhead{Type} &
\colhead{References} }
\startdata
ASASSN-14ae &  \dots &  11:08:40.11	 & $+$34:05:52.2 & TDE & \citet{holoien14b}, \citet{vanvelzen19b} \\
ASASSN-14li &  \dots & 12:48:15.230 & $+$17:46:26.44 & TDE & \citet{holoien16a}, \citet{brown17a}, \citet{vanvelzen19b} \\
ASASSN-15oi & \dots & 20:39:09.183 & $-$30:45:20.10 & TDE & \citet{holoien16b}, \citet{gezari17}, \citet{holoien18a} \\
ASASSN-17cv & AT2017bgt & 16:11:05.696 & $+$02:34:0.52 & AGN & \citet{trakhtenbrot19a} \\
ASASSN-18el & AT2018zf &  19:27:19.551 & $+$65:33:54.31 & AGN/TDE & \citet{trakhtenbrot19b}, \citet{ricci20} \\
ASASSN-18jd & AT2018bcb & 22:43:42.871 & $-$16:59:08.49 & AGN/TDE & \citet{neustadt20} \\
ASASSN-18pg & AT2018dyb & 16:10:58.774 & $-$60:55:23.16 & TDE & \citet[][]{leloudas19}, \citet[][]{Holoien20} \\
ASASSN-18ul & AT2018fyk & 22:50:16.090  & $-$44:51:53.50 & TDE & \citet[][]{wevers19} \\
ASASSN-18zj & AT2018hyz & 10:06:50.871 & $+$01:41:34.08 & TDE & \citet{vanvelzen20}, \citet{hung20a}, \citet{gomez20} \\
ASASSN-19bt & AT2019ahk & 07:00:11.546 & $-$66:02:24.14 & TDE & \citet{holoien19c} \\
ASASSN-19dj & AT2019azh & 08:13:16.945 & $+$22:38:54.03 & TDE & \citet{liu19}, \citet{vanvelzen20}, \citet{hinkle21a} \\
ATLAS18qqn & AT2018cow & 16:16:00.220 & $+$22:16:04.91 & Ambiguous & \citet{prentice18}, \citet{perley19} \\
ATLAS18way & AT2018hco & 01:07:33.635 & $+$23:28:34.28 & TDE & \citet{vanvelzen20} \\
ATLAS18yzs & AT2018iih & 17:28:03.930 & $+$30:41:31.42 & TDE & \citet{vanvelzen20} \\
ATLAS19qqu & AT2019mha & 16:16:27.799 & $+$56:25:56.29 & TDE & \citet{vanvelzen20} \\
Gaia19axp & AT2019brs & 14:27:46.400 & $+$29:30:38.27 & AGN &\citet{frederick20} \\
Gaia19bpt & AT2019ehz & 14:09:41.880 & $+$55:29:28.10 & TDE &  \citet{vanvelzen20} \\
iPTF15af & \dots & 08:48:28.120 & $+$22:03:33.58 & TDE & \citet{blagorodnova18} \\
iPTF16axa & \dots & 17:03:34.360 & $+$30:35:36.8 & TDE & \citet{hung17} \\
iPTF16fnl & AT2016fnl & 00:29:57.042 & $+$32:53:37.51 & TDE & \citet{blagorodnova17}, \citet{brown18} \\
OGLE16aaa & \dots & 01:07:20.880 & $-$64:16:20.70 & TDE & \citet{wyrzykowski17}, \citet{kajava20}, \citet{shu20} \\
OGLE17aaj & \dots & 01:56:24.930	& $-$71:04:15.70 & AGN & \citet{gromadzki19} \\
PS16dtm & AT2016ezh & 01:58:04.739  & $-$00:52:21.74 & TDE\tablenotemark{a} & \citet{blanchard17} \\
PS17dhz & AT2017eqx & 22:26:48.370 & $+$17:08:52.40 & TDE & \citet{nicholl19} \\
PS18kh & AT2018zr & 07:56:54.537 & $+$34:15:43.61 & TDE & \citet{holoien19b}, \citet{vanvelzen18} \\
ZTF18aahqkbt & AT2018bsi & 08:15:26.621	& $+$45:35:31.95 & TDE & \citet{vanvelzen20} \\
ZTF18aajupnt & AT2018dyk & 15:33:08.015	& $+$44:32:08.20 & LINER & \citet{frederick19} \\
ZTF18actaqdw & AT2018lni & 04:09:37.652	& $+$73:53:41.66 & TDE & \citet{vanvelzen20} \\
ZTF19aabbnzo & AT2018lna & 07:03:18.649	& $+$23:01:44.70& TDE & \citet{vanvelzen20} \\
ZTF19aaiqmgl & AT2019avd & 08:23:36.767	& $+$04:23:02.46 & AGN & \citet{frederick20} \\
ZTF19aakiwze & AT2019cho & 12:55:09.210 & $+$49:31:09.93 & TDE & \citet{vanvelzen20} \\
ZTF19aakswrb & AT2019bhf & 15:09:15.975	& $+$16:14:22.52 & TDE & \citet{vanvelzen20} \\
ZTF19aapreis & AT2019dsg & 20:57:02.974	 & $+$14:12:15.86 & TDE & \citet{vanvelzen20} \\
ZTF19aatubsj & AT2019fdr & 17:09:06.859	& $+$26:51:20.50 & TDE\tablenotemark{a} & \citet{frederick20} \\
ZTF19abhhjcc & AT2019meg & 18:45:16.180	& $+$44:26:19.21 & TDE & \citet{vanvelzen20} \\
ZTF19abidbya & AT2019lwu & 23:11:12.305 & $-$01:00:10.71 & TDE &  \citet{vanvelzen20} \\
ZTF19abvgxrq & AT2019pev & 04:29:22.720	& $+$00:37:07.50 & AGN &  \citet{frederick20} \\
ZTF19abzrhgq & AT2019qiz & 04:46:37.880	& $-$10:13:34.90 & TDE &  \citet{vanvelzen20}, \citet{nicholl20}, \citet{hung2020b} \\
\enddata 
\tablecomments{The 38 transients we re-analyze in this manuscript. TNS ID is the ID given for objects reported on the Transient Name Server. References include the discovery papers and papers using \swift data taken in 2015 or later. The type given reflects the classifications in the listed references. \textit{If using the revised photometry presented here, please cite both this paper and the original paper(s) in which \swift photometry was published.}}
\tablenotetext{a}{These sources have been interpreted as TDEs occuring in AGN host galaxies}
\label{tab:sample}
\end{deluxetable*}
 
\section{Host Galaxy SED Fits}

\label{sec:host_fits}
In order to accurately measure the UV and optical photometry of each transient, we must first subtract the emission of the host galaxy. Two of our sources, ASASSN-19bt and ATLAS18qqn have \swift images of the host galaxy during quiescence (see \citet{holoien19c} and \citet{perley19} respectively), from which we directly obtained host fluxes. In the case of ATLAS18qqn, we measured the flux at the location of the transient, offset from the host galaxy nucleus. For most of our sources, there was no archival \swift coverage of the host galaxy. For these sources, we fit archival multi-wavelength photometry of the host galaxy using the Fitting and Assessment of Synthetic Templates code \citep[\textsc{FAST};][]{kriek09} to obtain a spectral energy distribution (SED) of the host galaxy, from which we can estimate the UV flux.

\begin{deluxetable*}{ccccc}[htbp!]
%\tabletypesize{\textsize}
\tablecaption{Archival Host Photometry}
\tablehead{
\colhead{Object} &
\colhead{TNS ID} &
\colhead{Filter} &
\colhead{Magnitude} &
\colhead{Uncertainty}}
\startdata
\dots & \dots & \dots  & \dots  & \dots \\
ASASSN-19dj & AT2019azh & NUV  & 18.71  & 0.05 \\
ASASSN-19dj & AT2019azh & u(SDSS) &  16.80  & 0.10 \\
ASASSN-19dj & AT2019azh & g(SDSS) &  15.12 &  0.04 \\
ASASSN-19dj & AT2019azh & r(SDSS) &  14.59 &  0.03 \\
ASASSN-19dj & AT2019azh & i(SDSS) &  14.35 &  0.03 \\
ASASSN-19dj & AT2019azh & z(SDSS) &  14.13  & 0.03 \\
ASASSN-19dj & AT2019azh & J  & 13.94 &  0.04 \\
ASASSN-19dj & AT2019azh & H  & 13.99 &  0.09 \\
ASASSN-19dj & AT2019azh & K$_s$  & 14.34 &  0.05 \\
ASASSN-19dj & AT2019azh & W1 &  15.07 &  0.03 \\
ASASSN-19dj & AT2019azh & W2 &  15.70 &  0.03 \\
\dots & \dots & \dots  & \dots  & \dots \\
\enddata 
\tablecomments{Archival UV, optical, and infrared photometry used in the \textsc{FAST} SED fits for our objects. All magnitudes are presented in the AB system, using published conversions for systems naturally in the Vega system. For ASASSN-19bt and ATLAS18qqn, the UVOT magnitudes listed were used to subtract the \swift photometry and the other photometry was used for the host SED fit. The TDE ASASSN-19dj is shown here to illustrate the format, while the full table is available as an ancillary file.}
\label{tab:archival_phot}
\end{deluxetable*}

For objects without \swift images in quiescence, we used published host galaxy magnitudes to fit the host galaxy SED with \textsc{FAST} when available. For sources without literature magnitudes\footnote{The sources with host magnitudes in the literature are ASASSN-14ae, ASASSN-14li, ASASSN-15oi, ASASSN-18jd, ASASSN-18pg, ASASSN-18ul, ASASSN-18zj, ASASSN-19bt, ASASSN-19dj, iPTF16fnl, OGLE17aaj, and PS18kh}  we obtained $JHK_S$ images from the Two Micron All-Sky Survey \citep[2MASS;][]{skrutskie06} for near infrared (NIR) constraints and $ugriz$ or $grizY$ images from the from Sloan Digital Sky Survey (SDSS) Data Release 16 \citep{ahumada20} or Pan-STARRS \citep[][]{chambers16} for optical constraints. We then measured aperture magnitudes of the host galaxy in the 2MASS and SDSS/Pan-STARRS data using the same aperture size as was used for the follow-up photometry (see Section~\ref{sec:swift_reductions}), using nearby stars to calibrate the galaxy magnitudes. The TDEs ASASSN-19bt and OGLE16aaa were too far south to be observed by either SDSS or Pan-STARRS, so we obtained catalog magnitudes from the AAVSO Photometric All-Sky Survey \citep[APASS;][]{henden15} and the Dark Energy Survey \citep{abbott18}, respectively. We additionally obtained UV magnitudes from the Galaxy Evolution Explorer \citep[GALEX;][]{martin05} All-sky Imaging Survey (AIS) catalog and $W1$ and $W2$ magnitudes from the Wide-field Infrared Survey Explorer \citep[WISE;][]{wright10} AllWISE catalog for all hosts in our sample. The archival photometry is shown in Table \ref{tab:archival_phot}. 

We then fit this archival host-galaxy photometry using \textsc{FAST} and assuming a \citet{cardelli88} extinction law with $\text{R}_{\text{V}} = 3.1$ and Galactic extinction at the coordinates of the host galaxy \citep{schlafly11}, a Salpeter IMF \citep{salpeter55}, an exponentially declining star-formation rate, and the \citet{bruzual03} stellar population models. We estimated the host flux in the UVOT filters for each \textsc{FAST} iteration by convolving the best-fit host SED from \textsc{FAST} with the filter response curve for each filter, obtained from the Spanish Virtual Observatory (SVO) Filter Profile Service \citep{rodrigo12}. In addition to the UVOT filters, we used the Bessel filter responses \citep{bessel90} to obtain Johnson-Cousins magnitudes. To estimate the uncertainties on the estimated host-galaxy fluxes, we performed a Monte Carlo sampling by perturbing the archival host fluxes assuming Gaussian errors and running 1000 different \textsc{FAST} iterations for each host galaxy. We took the median value as the magnitude and calculated $1\sigma$ errors by taking the difference between the 16th and 84th percentile values from the median and taking the larger value as the error. We then subtracted these synthetic host fluxes from the \swift photometry. In most cases, these synthetic magnitudes are well-constrained, but for host galaxies without GALEX magnitudes, such as the hosts of ASASSN-18pg, PS17dhz and ZTF19abidbya, the UV synthetic magnitudes often have large uncertainties as the star formation rates (SFRs), and thus UV emission, are poorly constrained. The synthetic magnitudes computed for each object, spanning from GALEX $FUV$ to 2MASS $K_s$ are shown in Table \ref{tab:synth_phot}. In general, the archival and synthetic magnitudes agree within the uncertainties indicating a reasonable fit. For the objects with synthetic host magnitudes in the literature, such as PS18kh and ASASSN-18ul, our host values are largely consistent within the uncertainties. Any discrepancies are likely due to fitting different archival photometry and/or different choices made when fitting the host SEDs.

Because the TDE ASASSN-19bt \citep{holoien19c} has \swift UVOT images with no transient source, we are able to test the accuracy of our synthetic \swift magnitudes. We fit GALEX $NUV$, APASS $gri$, 2MASS $JHK_s$, WISE $W1$ and $W2$ photometry of the host galaxy, excluding the UVOT data, and then computed synthetic UVOT magnitudes. We find that for each of the six UVOT bands, the measured and synthetic photometry are consistent given the uncertainties. If we repeat this process without the GALEX $NUV$ constraint, the differences are larger but the models are still consistent with the data given the larger uncertainties.

\begin{deluxetable*}{cccccc}[htbp!]
%\tabletypesize{\textsize}
\tablecaption{Synthetic Host-Galaxy Magnitudes}
\tablehead{
\colhead{Object} &
\colhead{TNS ID} &
\colhead{Filter} &
\colhead{Magnitude} &
\colhead{Uncertainty}}
\startdata
\dots & \dots & \dots  & \dots  & \dots  \\
ASASSN-19dj & AT2019azh &  FUV(GALEX) & 20.449 & 0.863\\
ASASSN-19dj & AT2019azh &  NUV(GALEX) & 18.768 & 0.125\\
ASASSN-19dj & AT2019azh &  UVW2(UVOT) & 19.320 & 0.238\\
ASASSN-19dj & AT2019azh &  UVM2(UVOT) & 18.827 & 0.127\\
ASASSN-19dj & AT2019azh &  UVW1(UVOT)& 18.196 & 0.108\\
ASASSN-19dj & AT2019azh &  U(UVOT) &  16.721 & 0.048\\
ASASSN-19dj & AT2019azh &  B(UVOT) & 15.445 & 0.079\\
ASASSN-19dj & AT2019azh &  V(UVOT) & 14.884 & 0.040\\
ASASSN-19dj & AT2019azh &  U(J-C) & 16.574 & 0.055\\
ASASSN-19dj & AT2019azh &  B(J-C) & 15.439 & 0.076\\
ASASSN-19dj & AT2019azh &  V(J-C) & 14.830 & 0.039\\
ASASSN-19dj & AT2019azh &  R(J-C) & 14.580 & 0.022\\
ASASSN-19dj & AT2019azh &  I(J-C) & 14.331 & 0.022\\
ASASSN-19dj & AT2019azh &  u(SDSS) & 16.623 & 0.049\\
ASASSN-19dj & AT2019azh &  g(SDSS) & 15.203 & 0.064\\
ASASSN-19dj & AT2019azh &  r(SDSS) & 14.630 & 0.026\\
ASASSN-19dj & AT2019azh &  i(SDSS) & 14.405 & 0.019\\
ASASSN-19dj & AT2019azh  & z(SDSS) & 14.211 & 0.029\\
ASASSN-19dj & AT2019azh  & J(2MASS) & 13.995 & 0.041\\
ASASSN-19dj & AT2019azh  & H(2MASS) & 13.798 & 0.050\\
ASASSN-19dj & AT2019azh  & K$_s$(2MASS) & 13.989 & 0.049\\
\dots & \dots & \dots  & \dots  & \dots & \\
\enddata 
\tablecomments{Synthetic host photometry computed from the Monte Carlo sampling of host galaxy SED fits with \textsc{FAST}. We used the Bessel filter responses \citep{bessel90} for our Johnson-Cousins synthetic magnitude calculations. All magnitudes are presented in the AB system, using published conversions for systems naturally in the Vega system. Note that the synthetic photometry listed in this table for ATLAS18qqn represents the entire host galaxy, not the region where the transient occurred and thus the UVOT magnitudes are significantly brighter. The values used for the subtraction of ATLAS18qqn are measured at the region of the transient from \swift images. The TDE ASASSN-19dj is shown here to illustrate the format and the full table is available as an ancillary file.}
\label{tab:synth_phot}
\end{deluxetable*}

In addition to providing synthetic photometry for host flux subtraction, the \textsc{FAST} models constrain the age of the stellar population, the stellar mass, and the star formation rate of the host. Uncertainties on the host properties are computed in the same fashion as the synthetic magnitudes. In some cases our reported one-sided uncertainties are zero, which is a consequence of the grid spacing used in the \textsc{FAST} fitting procedure. In such cases, the median and either 16th or 84th percentile values are identical due to the discrete spacing of the grid in that parameter. This is most notable in the stellar population ages where the grid spacing was log(age) = 0.05. The SED fits to the host galaxies of ZTF19aabbnzo and iPTF6axa are particularly poorly constrained, with no $1\sigma$ limit on the SFR given the sampling. If we instead employ more relaxed limits to obtain a constraint, the SFR for the host galaxy of ZTF19aabbnzo has a $3\sigma$ upper limit of log[SFR (M$\odot$ yr$^{-1}$)] $< -1.86$ and the host galaxy of iPTF16axa has a $4\sigma$ upper limit of log[SFR (M$\odot$ yr$^{-1}$)] $< -2.90$.

Table \ref{tab:host_props} provides these host parameters for each of our host galaxies. \textsc{FAST} only fits stellar population synthesis models, so the fits for the galaxies known to host AGN have not taken into account a non-stellar component. Additionally for some of the larger galaxies, the default 5\farcs{0} radius used to match the \swift photometry does not encapsulate the full host galaxy. Finally, as expected, many of the TDE host galaxies are relatively low mass, consistent with hosting SMBHs less massive than $\sim 10^7$ M$\odot$ \citep{ vanvelzen18,wevers19,mockler19}.

\begin{deluxetable*}{ccccccc}[htbp!]
%\tabletypesize{\textsize}
\tablecaption{Host-Galaxy Properties}
\tablehead{
\colhead{Object} &
\colhead{TNS ID} &
\colhead{Redshift} &
\colhead{A$_{V}$} &
\colhead{log[Age (yr)]} &
\colhead{log[Mass (M$\odot$)]} &
\colhead{log[SFR (M$\odot$ yr$^{-1}$)]}
}
\startdata
ASASSN-14ae & \dots & 0.0436 & 0.048 & $9.25_{-0.05}^{+0.00}$ & $9.78_{-0.03}^{+0.01}$ &  $<-5.03$ \\
ASASSN-14li  & \dots & 0.0206 & 0.070 & $9.05_{-0.00}^{+0.05}$  & $9.44_{-0.01}^{+0.01}$ &  $-2.23_{-0.79}^{+0.17}$ \\ 
ASASSN-15oi  & \dots & 0.0479 & 0.185 & $9.35_{-0.05}^{+0.05}$  & $9.99_{-0.04}^{+0.03}$ &  $<-3.60$ \\ 
ASASSN-17cv  & AT2017bgt &  0.0640 & 0.213 & $9.95_{-0.15}^{+0.00}$  & $11.06_{-0.07}^{+0.01}$ &  $0.32_{-0.08}^{+0.02}$ \\
ASASSN-18el  & AT2018zf & 0.0190 & 0.236 & $9.65_{-0.05}^{+0.10}$  & $9.59_{-0.06}^{+0.02}$ &  $-0.81_{-0.06}^{+0.03}$ \\ 
ASASSN-18jd  & AT2018bcb & 0.1192 & 0.098 & $9.70_{-0.10}^{+0.15}$  & $11.07_{-0.09}^{+0.08}$ &  $-0.41_{-0.02}^{+0.26}$ \\ 
ASASSN-18pg  & AT2018dyb & 0.0179 & 0.624 &  $9.90_{-0.20}^{+0.10}$  & $10.16_{-0.11}^{+0.05}$ &  $<-1.13$ \\ 
ASASSN-18ul  & AT2018fyk & 0.0590 & 0.037 & $9.70_{-0.00}^{+0.10}$  & $10.64_{-0.02}^{+0.05}$ &  $-1.50_{-1.99}^{+0.40}$ \\
ASASSN-18zj  & AT2018hyz & 0.0457 & 0.094 & $9.10_{-0.00}^{+0.10}$  & $9.71_{-0.04}^{+0.04}$ &  $<-1.85$ \\
ASASSN-19bt  & AT2019ahk & 0.0262 & 0.336 & $9.85_{-0.20}^{+0.15}$  & $10.29_{-0.12}^{+0.08}$ &  $-0.60_{-0.03}^{+0.04}$ \\ 
ASASSN-19dj  & AT2019azh & 0.0223 & 0.122 & $9.10_{-0.05}^{+0.05}$  & $9.95_{-0.02}^{+0.03}$ &  $-1.56_{-0.17}^{+0.42}$ \\ 
ATLAS18qqn  & AT2018cow &  0.0141 & 0.238 & $9.95_{-0.10}^{+0.05}$  & $9.81_{-0.06}^{+0.01}$ &  $-0.69_{-0.03}^{+0.01}$ \\ 
ATLAS18way  & AT2018hco & 0.0880 & 0.109 & $9.15_{-0.10}^{+0.25}$  & $9.69_{-0.06}^{+0.17}$ &  $<-0.95$ \\ 
ATLAS18yzs  & AT2018iih & 0.2120 & 0.135 & $9.10_{-0.00}^{+0.40}$  & $10.56_{-0.06}^{+0.26}$ &  $<-2.07$ \\
ATLAS19qqu  & AT2019mha & 0.1480 & 0.022 & $8.95_{-0.05}^{+0.20}$  & $9.75_{-0.07}^{+0.08}$ &  $<-1.63$ \\ 
Gaia19axp  & AT2019brs & 0.3736 & 0.043 & $9.45_{-0.85}^{+0.45}$  & $11.08_{-0.52}^{+0.35}$ &  $0.57_{-2.15}^{+0.05}$ \\ 
Gaia19bpt  & AT2019ehz & 0.0740 & 0.048 & $9.10_{-0.10}^{+0.15}$  & $9.37_{-0.05}^{+0.13}$ &  $<-2.07$ \\ 
iPTF15af & \dots & 0.0790 & 0.093 & $9.30_{-0.00}^{+0.10}$ & $10.07_{-0.04}^{+0.08}$ & $<-8.59$ \\
iPTF16axa & \dots & 0.1080 & 0.124 & $9.00_{-0.00}^{+0.00}$ & $10.60_{-0.02}^{+0.02}$ & $-99.00_{-0.00}^{+0.00}$ \\
iPTF16fnl  & AT2016fnl & 0.0163 & 0.226 & $9.25_{-0.00}^{+0.05}$  & $9.49_{-0.00}^{+0.04}$ &  $-1.73_{-0.01}^{+0.15}$ \\ 
OGLE16aaa  & \dots & 0.1655 & 0.057 & $9.20_{-0.00}^{+0.05}$  & $10.54_{-0.01}^{+0.02}$ &  $-0.46_{-0.01}^{+0.28}$ \\ 
OGLE17aaj  & \dots  & 0.1160 & 0.077 & $9.70_{-0.85}^{+0.20}$ & $10.41_{-0.33}^{+0.13}$ & $<0.13$\\
PS16dtm  & AT2016ezh &  0.0804 & 0.070 & $9.50_{-0.20}^{+0.35}$  & $9.93_{-0.14}^{+0.20}$ &  $-0.55_{-0.05}^{+0.03}$ \\ 
PS17dhz  & AT2017eqx & 0.1089 & 0.175 & $9.30_{-1.05}^{+0.70}$  & $9.12_{-0.76}^{+0.38}$ &  $<-0.13$ \\ 
PS18kh  & AT2018zr &  0.0710 & 0.128 & $9.50_{-0.05}^{+0.00}$  & $9.97_{-0.04}^{+0.02}$ &  $<-2.94$ \\ 
ZTF18aahqkbt  & AT2018bsi & 0.0510 & 0.170 & $9.80_{-0.10}^{+0.00}$  & $10.72_{-0.08}^{+0.01}$ &  $-0.86_{-0.01}^{+0.02}$ \\ 
ZTF18aajupnt &	AT2018dyk & 0.0367 & 0.055 & $9.85_{-0.00}^{+0.10}$  & $11.00_{-0.01}^{+0.07}$ &  $0.05_{-0.02}^{+0.01}$ \\
ZTF18actaqdw  & AT2018lni & 0.1380 & 0.661 & $10.00_{-0.05}^{+0.00}$  & $10.23_{-0.05}^{+0.06}$ &  $-0.28_{-0.14}^{+0.10}$ \\ 
ZTF19aabbnzo  & AT2018lna & 0.0910 & 0.130 & $9.00_{-0.00}^{+0.00}$  & $9.76_{-0.02}^{+0.03}$ &  $-99.00_{-0.00}^{+0.00}$ \\ 
ZTF19aaiqmgl  & AT2019avd & 0.0296 & 0.072 & $9.90_{-0.10}^{+0.05}$  & $10.42_{-0.06}^{+0.02}$ &  $-0.77_{-0.03}^{+0.04}$ \\
ZTF19aakiwze  & AT2019cho & 0.1930 & 0.038 & $9.20_{-0.30}^{+0.60}$  & $9.82_{-0.18}^{+0.31}$ &  $<-0.42$ \\
ZTF19aakswrb  & AT2019bhf & 0.1206 & 0.068 & $9.50_{-0.20}^{+0.30}$  & $10.34_{-0.14}^{+0.20}$ &  $-0.71_{-0.41}^{+0.16}$ \\ 
ZTF19aapreis  & AT2019dsg & 0.0512 & 0.280 & $9.85_{-0.05}^{+0.05}$  & $10.57_{-0.02}^{+0.02}$ &  $-0.75_{-0.26}^{+0.12}$ \\
ZTF19aatubsj  & AT2019fdr & 0.2666 & 0.145 & $9.45_{-0.20}^{+0.25}$  & $10.93_{-0.13}^{+0.17}$ &  $0.22_{-0.17}^{+0.18}$ \\ 
ZTF19abhhjcc  & AT2019meg & 0.1520 & 0.153 & $9.93_{-0.23}^{+0.07}$  & $10.16_{-0.11}^{+0.07}$ &  $-0.79_{-0.26}^{+0.10}$ \\
ZTF19abidbya  & AT2019lwu & 0.1170 & 0.101 & $10.00_{-0.90}^{+0.00}$  & $9.98_{-0.38}^{+0.31}$ &  $<-0.20$ \\ 
ZTF19abvgxrq  & AT2019pev & 0.0970 & 0.225 & $9.05_{-0.05}^{+0.10}$  & $10.33_{-0.03}^{+0.05}$ &  $0.09_{-0.15}^{+0.07}$ \\ 
ZTF19abzrhgq  & AT2019qiz & 0.0151 & 0.302 & $9.50_{-0.00}^{+0.15}$  & $9.89_{-0.02}^{+0.08}$ &  $<-2.28$ \\ 
\enddata 
\tablecomments{Host-galaxy properties computed from the \textsc{FAST} SED models in addition to the host-galaxy redshift and Galactic visual extinction \citep{schlafly11}. It is important to note that the radii used to measure the host photometry were chosen to match the \swift aperture radius and therefore for some objects do not encompass the entire host galaxy.}
\label{tab:host_props}
\end{deluxetable*}

\section{\swift UVOT Reductions}
\label{sec:swift_reductions}
The UVOT has six typically used filters for photometric follow-up programs \citep{poole08}: $V$ (5425.3 \AA), $B$ (4349.6 \AA), $U$ (3467.1 \AA), $UVW1$ (2580.8 \AA), $UVM2$ (2246.4 \AA), and $UVW2$ (2054.6 \AA). The wavelengths quoted here are the pivot wavelengths calculated by the SVO Filter Profile Service \citep{rodrigo12}, which we use throughout the remainder of this work. 

Most epochs of UVOT data include multiple observations in each filter. We separately combined the images in each filter for each unique observation identification number using the HEASoft {\tt uvotimsum} package. We then used the {\tt uvotsource} package to extract source counts in a region centered on the position of the transient and background counts using a source-free region with radius of $\sim$30-40\farcs{}. Our default source radius was 5\farcs{0} to minimize UVOT aperture corrections. We then converted the UVOT count rates into fluxes and magnitudes using the most recent calibrations \citep{poole08, breeveld10}. For each UVOT image, we confirmed that the source did not lie on a region of the detector with known sensitivity issues\footnote{\url{https://swift.gsfc.nasa.gov/analysis/uvot_digest/sss_check.html}} (also see the Appendix of \citealt[][]{edelson15}).

As the UVOT uses unique $B$ and $V$ filters, we converted the UVOT $B$ and $V$ data into the Johnson-Cousins system using color corrections\footnote{\url{https://heasarc.gsfc.nasa.gov/docs/heasarc/caldb/swift/docs/uvot/uvot_caldb_coltrans_02b.pdf}}. For these filters, we used pivot wavelengths of $V$ (5477.7 \AA) and $B$ (4371.1 \AA), corresponding to the Bessel filter responses used in the synthetic magnitude calculations. Table \ref{tab:raw_swift} provides the \swift photometry in both magnitude and flux density without host subtraction or extinction correction.

\begin{deluxetable*}{cccccccc}[htbp!]
%\tabletypesize{\textsize}
\tablecaption{Unsubtracted \swift Photometry}
\tablehead{
\colhead{Object} &
\colhead{TNS ID} &
\colhead{MJD} &
\colhead{Filter} &
\colhead{Magnitude} &
\colhead{Uncertainty} &
\colhead{Flux Density} &
\colhead{Uncertainty}\\
&
&
&
&
& &
\colhead{(erg s$^{-1}$ cm$^{-2}$ \AA$^{-1}$)} &
\colhead{erg s$^{-1}$ cm$^{-2}$ \AA$^{-1}$)}}
\startdata
\dots & \dots & \dots  & \dots  & \dots & \dots & \dots  & \dots \\
ASASSN-19dj   & AT2019azh  &  58544.762  &  V    &	14.50  &  0.03   & 5.83E-15 & 1.77E-15 \\
ASASSN-19dj   & AT2019azh  &  58553.457   & V    &	14.46  &  0.04  &  6.10E-15 & 0.24E-15\\
\dots & \dots & \dots  & \dots  & \dots & \dots & \dots & \dots  \\
ASASSN-19dj  &  AT2019azh   & 58544.758  &  B    &	14.68  &  0.04   & 7.71E-15 & 0.32E-15\\
ASASSN-19dj  & AT2019azh &  58553.454 &  B   &	14.53 &  0.04  & 8.85E-15  & 0.37E-15\\
\dots & \dots & \dots  & \dots  & \dots & \dots & \dots  & \dots \\
ASASSN-19dj   & AT2019azh   & 58544.757  &  U    &	15.00   & 0.03   & 9.04E-15 & 0.25E-15\\ 
ASASSN-19dj  & AT2019azh &  58553.453 &  U   &	14.88 &  0.04  & 1.01E-14 &  0.04E-14\\
\dots & \dots & \dots  & \dots  & \dots & \dots & \dots  & \dots \\
ASASSN-19dj   & AT2019azh   & 58544.755  &  UVW1    &	15.00   & 0.04  &  1.63E-14 & 0.06E-14\\ 
ASASSN-19dj &  AT2019azh &  58553.451  & UVW1  & 	14.92 &  0.04 &  1.76E-14 &  0.06E-14\\
\dots & \dots & \dots  & \dots  & \dots & \dots & \dots  & \dots \\
ASASSN-19dj  &  AT2019azh   & 58544.763  &  UVM2    &	14.92   & 0.04   & 2.32E-14 &  0.06E-14\\ 
ASASSN-19dj   & AT2019azh  &  58553.458  & UVM2    &	14.88   & 0.04   & 2.41E-14  & 0.09E-14\\
\dots & \dots & \dots  & \dots  & \dots & \dots & \dots & \dots  \\
ASASSN-19dj  &  AT2019azh   & 58544.759  &  UVW2   & 	14.75  &  0.04   & 3.24E-14 & 0.12E-14\\ 
ASASSN-19dj  & AT2019azh  & 58553.454  & UVW2   &	14.67  & 0.04  & 3.49E-14  & 0.13E-14\\
\dots & \dots & \dots  & \dots  & \dots & \dots & \dots  & \dots \\
\enddata 
\tablecomments{\swift photometry of the transients without the host flux subtracted and with no correction for Galactic extinction. The $BV$ photometry has been converted to the Johnson-Cousins system using the color-corrections described in the text. All magnitudes are presented in the AB system, using published conversions for systems naturally in the Vega system. The data for each source are grouped by filter and sorted by increasing MJD. The TDE ASASSN-19dj is shown here to illustrate the format and the full table is available as an ancillary file.}
\label{tab:raw_swift}
\end{deluxetable*}

After computing the raw \swift photometry, and correcting the $BV$ data to the Johnson-Cousins system, we corrected each epoch of UVOT photometry for Galactic extinction \citep{schlafly11} (see Table \ref{tab:archival_phot}) and removed the host contamination by subtracting the corresponding host flux in each filter. To compute the uncertainties, we added the uncertainty in the \swift photometry and the uncertainty in the host flux in that filter in quadrature. These results are provided in Table \ref{tab:sutracted_swift}. Where the transient flux was less than a 3$\sigma$ detection, we give a 3$\sigma$ upper limit on the transient magnitude.

\begin{deluxetable*}{cccccccc}[htbp!]
%\tabletypesize{\textsize}
\tablecaption{Host-Subtracted \swift Photometry}
\tablehead{
\colhead{Object} &
\colhead{TNS ID} &
\colhead{MJD} &
\colhead{Filter} &
\colhead{Magnitude} &
\colhead{Uncertainty} &
\colhead{Flux Density} &
\colhead{Uncertainty}\\
&
&
&
&
& &
\colhead{(erg s$^{-1}$ cm$^{-2}$ \AA$^{-1}$)} &
\colhead{erg s$^{-1}$ cm$^{-2}$ \AA$^{-1}$)}}
\startdata
\dots & \dots & \dots  & \dots  & \dots & \dots & \dots  & \dots \\
ASASSN-19dj  &  AT2019azh   & 58544.762  &  V    &	15.85  &  0.17  &  1.66E-15  &  0.26E-15\\
ASASSN-19dj  &  AT2019azh   & 58553.457  &  V    &	15.67  &  0.17  &  1.95E-15  &  0.31E-15\\
\dots & \dots & \dots  & \dots  & \dots & \dots & \dots  & \dots \\
ASASSN-19dj  &  AT2019azh  &  58544.758  &  B   & 	15.27  &  0.12  &  4.46E-15  &  0.48E-15\\
ASASSN-19dj  &  AT2019azh  &  58553.454  &  B    &	14.98  &  0.10  &  5.77E-15  &  0.52E-15\\
\dots & \dots & \dots  & \dots  & \dots & \dots & \dots & \dots  \\ 
ASASSN-19dj  &  AT2019azh   & 58544.757  &  U    &	15.05  &  0.04   & 8.61E-15  &  0.31E-15\\
ASASSN-19dj  &  AT2019azh   & 58553.453  &  U    &	14.90  &  0.05  &  9.87E-15  &  0.45E-15\\
\dots & \dots & \dots  & \dots  & \dots & \dots & \dots & \dots  \\
ASASSN-19dj  &  AT2019azh   & 58544.755  &  UVW1  &  	14.79   & 0.04  & 1.98E-14  &  0.08E-14\\
ASASSN-19dj  &  AT2019azh   & 58553.451  &  UVW1  &  	14.71   & 0.04  &  2.13E-14  & 0.08E-14\\
\dots & \dots & \dots  & \dots  & \dots & \dots & \dots  & \dots \\ 
ASASSN-19dj   & AT2019azh   & 58544.763  &  UVM2  &  	14.58  &  0.04   & 3.17E-14  &  0.12E-14\\
ASASSN-19dj   & AT2019azh  &  58553.458  &  UVM2  &  	14.54  &  0.04   & 3.30E-14  &  0.13E-14\\
\dots & \dots & \dots  & \dots  & \dots & \dots & \dots  & \dots \\
ASASSN-19dj  &  AT2019azh  &  58544.759  &  UVW2  &  	14.40  &  0.04   & 4.48E-14  &  0.17E-14\\
ASASSN-19dj  & AT2019azh   & 58553.454  &  UVW2   & 	14.32  &  0.04   & 4.83E-14  &  0.18E-14\\
\dots & \dots & \dots  & \dots  & \dots & \dots & \dots  & \dots \\
\enddata 
\tablecomments{\swift photometry of the transients with the host flux subtracted corrected for Galactic extinction. The uncertainties incorporate both the error on the photometry and from the host SED fits. For epochs where the transient flux was less than a 3$\sigma$ detection, the magnitude column shows a 3$\sigma$ upper limit on the transient magnitude. All magnitudes are presented in the AB system, using published conversions for systems naturally in the Vega system. The data for each source are grouped by filter and sorted by increasing MJD. The TDE ASASSN-19dj is shown here to illustrate the format and the full table is available as an ancillary file.}
\label{tab:sutracted_swift}
\end{deluxetable*}

\section{Blackbody Fits} \label{sec:bb_fits}

The host-subtracted UV/optical SEDs of TDEs \citep[e.g.,][]{holoien14b, holoien16a} and some AGN flares \citep[e.g.,][]{neustadt20} are well-fit as blackbodies. While in AGN the geometry of the emitting region is likely non-spherical and the emission is at least partly non-thermal, a simple blackbody fit should provide a reasonable estimate of the size and luminosity of the optically thick, continuum-emitting region. Therefore, we include blackbody fits for each of the objects in our sample for completeness. For each epoch of host-subtracted UV photometry, we used Markov Chain Monte Carlo (MCMC) methods to fit a blackbody model, as used in \citet[]{holoien14b, holoien16a}. The date listed for each epoch is the mean MJD of the data used in the fit. Unlike our $3\sigma$ detection limit for our reported \swift magnitudes, we employ a more liberal 2$\sigma$ detection threshold for our blackbody fits as the models are fit in flux space and this allows for marginal detections of the transient at late times. We do not include blackbody fits for the TDEs ZTF18actaqdw, ZTF19aabbnzo, and ZTF19aakiwze, as their \swift light curves only have coverage in a single filter and we cannot constrain the temperature.

\begin{deluxetable*}{cccccccccccc}[htbp!]
\tablecaption{Blackbody Fits}
\tablehead{
\colhead{Object} &
\colhead{TNS ID} &
\colhead{MJD} &
\colhead{log(L)} &
\colhead{dlog(L$_{l}$)} &
\colhead{dlog(L$_{u}$)} &
\colhead{log(R)} &
\colhead{dlog(R$_{l}$)} &
\colhead{dlog(R$_{u}$)} &
\colhead{log(T)} &
\colhead{dlog(T$_{l}$)} &
\colhead{dlog(T$_{u}$)}\\
\colhead{} &
\colhead{} &
\colhead{} &
\multicolumn{3}{c}{log([erg s$^{-1}$])} &
\multicolumn{3}{c}{log([cm])} &
\multicolumn{3}{c}{log([K])} 
}
\startdata
\dots & \dots & \dots  & \dots  & \dots & \dots & \dots & \dots & \dots & \dots & \dots & \dots \\
ASASSN-19dj & AT2019azh & 58544.76 & 44.45 & 0.07 & 0.08 & 14.64 & 0.04 & 0.04 & 4.58 & 0.03 & 0.04 \\
ASASSN-19dj & AT2019azh & 58553.45 & 44.36 & 0.06 & 0.07 & 14.72 & 0.04 & 0.04 & 4.52 & 0.03 & 0.04 \\
ASASSN-19dj & AT2019azh & 58556.11 & 44.41 & 0.05 & 0.06 & 14.77 & 0.04 & 0.03 & 4.50 & 0.03 & 0.03 \\
ASASSN-19dj & AT2019azh & 58562.95 & 44.28 & 0.04 & 0.04 & 14.83 & 0.03 & 0.03 & 4.44 & 0.02 & 0.02 \\
ASASSN-19dj & AT2019azh & 58565.94 & 44.32 & 0.04 & 0.05 & 14.79 & 0.03 & 0.03 & 4.47 & 0.02 & 0.03 \\
ASASSN-19dj & AT2019azh & 58568.23 & 44.21 & 0.03 & 0.03 & 14.82 & 0.02 & 0.02 & 4.43 & 0.02 & 0.02 \\
ASASSN-19dj & AT2019azh & 58574.79 & 44.24 & 0.04 & 0.04 & 14.80 & 0.03 & 0.03 & 4.44 & 0.02 & 0.02 \\
ASASSN-19dj & AT2019azh & 58577.10 & 44.24 & 0.04 & 0.04 & 14.79 & 0.03 & 0.02 & 4.45 & 0.02 & 0.02 \\
ASASSN-19dj & AT2019azh & 58580.62 & 44.25 & 0.04 & 0.05 & 14.75 & 0.03 & 0.03 & 4.47 & 0.02 & 0.03 \\
\dots & \dots & \dots  & \dots  & \dots & \dots & \dots & \dots & \dots & \dots & \dots & \dots \\
\enddata 
\tablecomments{Bolometric luminosity, effective radius, and temperature estimated from the blackbody fits to the host-subtracted and extinction-corrected \swift data. 
%All uncertainties are given in log space. 
The TDE ASASSN-19dj is shown here to illustrate the format and the full table is available as an ancillary file.}
\label{tab:BB_fits}
\end{deluxetable*}

\begin{figure*}
\centering
 \includegraphics[width=1.0\textwidth]{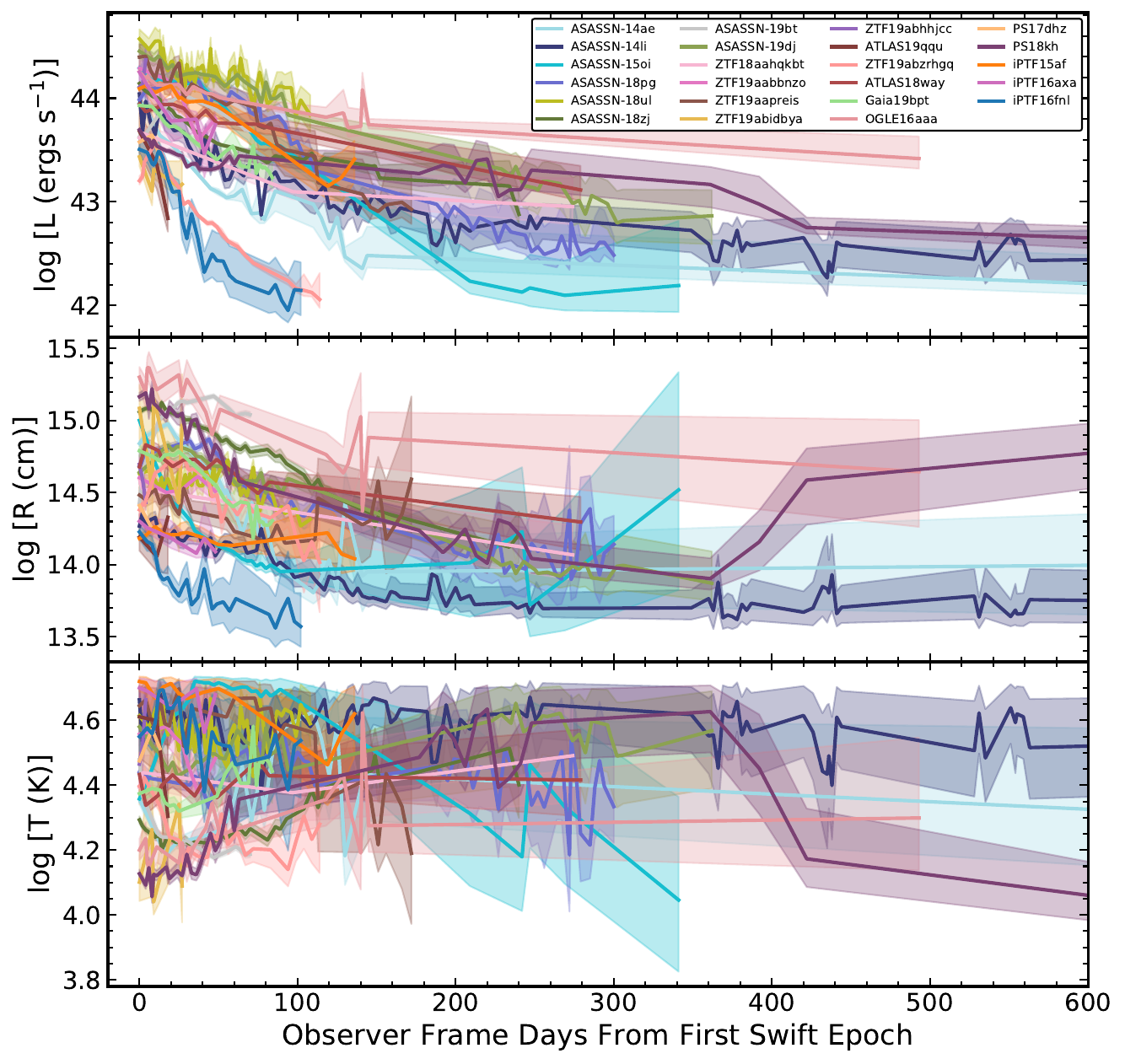}\hfill
 \caption{Evolution of the UV/optical blackbody luminosity (top panel), effective radius (middle panel), and temperature (bottom panel) for the TDEs analyzed in this work. The shading corresponds to the uncertainty. Time is in observer-frame days relative to the earliest \swift epoch. We have not shown the very-late time blackbody properties for ASASSN-14ae, ASASSN-14li, and PS18kh, which are included in Table \ref{tab:BB_fits}, to allow the evolution of the other TDEs to be seen more clearly.}
 \label{fig:TDE_BBs}
\end{figure*}

The \swift UVOT calibration correction only affected the UV filters, making them fainter than previously measured. This caused most objects to become cooler and therefore $\sim$15\% - 30\% less luminous than estimated from earlier reductions of \swift data. The evolution of blackbody parameters for the TDEs in this sample are shown in Figure \ref{fig:TDE_BBs}. Even with corrections to the \swift UV data, all of the TDEs are hot, with temperatures of $\sim 15,000 - 50,000$ K. The temperatures are roughly constant with time, although some objects show trends in their temperatures, both consistent with previous results \citep[e.g.,][]{hinkle20a, vanvelzen20}. As noted in \citet{hinkle20a}, the more luminous TDEs appear to decay more slowly than their less luminous counterparts (see top panel of Figure \ref{fig:TDE_BBs}).

\begin{figure*}
\centering
 \includegraphics[width=1.0\textwidth]{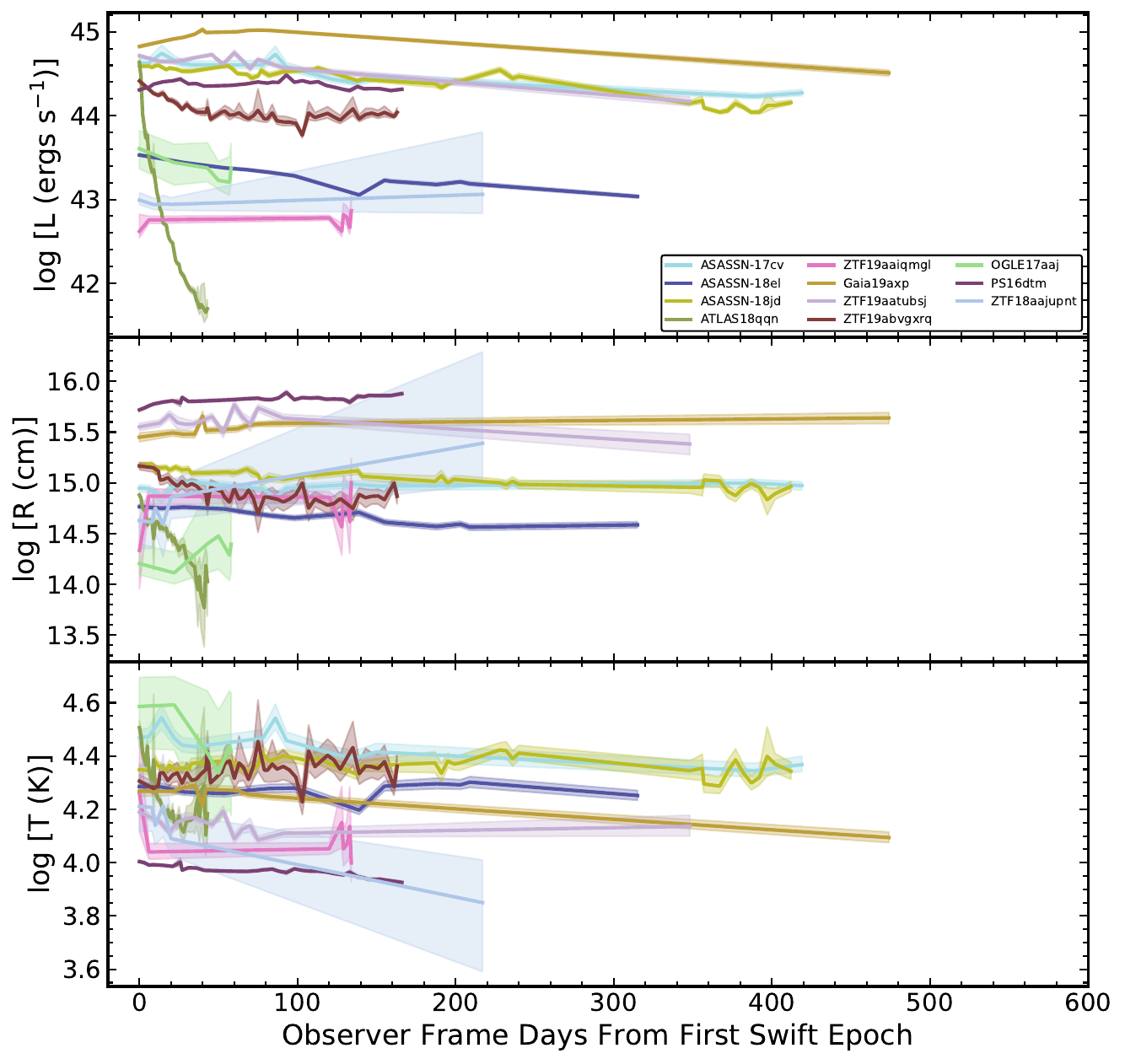}\hfill
 \caption{Evolution of the UV/optical blackbody luminosity (top panel), effective radius (middle panel), and temperature (bottom panel) for the non-TDE transients or sources interpreted as TDEs in AGN host galaxies. The shading corresponds to the uncertainty. Time is in observer-frame days relative to the earliest \swift epoch.}
 \label{fig:other_BBs}
\end{figure*}

Figure \ref{fig:other_BBs} shows the blackbody fits for the other nuclear outbursts. The luminosity range of these objects is much larger than for the TDEs because they span several source classes. The blackbody temperatures are still hot, consistent with the lower temperature range of TDEs. For many objects, the evolution in luminosity, radius, and temperature is much slower than for the TDEs, potentially due to these AGN hosting more massive SMBHs than the TDE hosts, although there is significant scatter in the various estimates of SMBH mass for some of these sources \citep[e.g.,][]{frederick20}. Additionally, unlike the TDEs, there seems to be no trend between peak luminosity and decline rate, consistent with the analysis of comparison objects in \citet{hinkle20a}.

\section{Peak-luminosity/Decline-rate Relationship} \label{sec:L40}
In Figure \ref{fig:TDE_BBs}, the most luminous TDEs appear to have flatter slopes near peak, and decay more slowly than the less luminous TDEs, consistent with the relationship presented in \citep{hinkle20a}. Given the importance of UV photometry to the bolometric UV/optical lightcurves on which this relationship is based, we re-analyzed this relationship with our updated \swift data. Similar to \citet{hinkle20a}, we have bolometrically corrected epochs without \swift UV data using nearby \swift epochs. Because the process of bolometrically correcting ground-based data involves heterogeneous data, we do not include the results of these bolometric corrections in Figure \ref{fig:TDE_BBs}, rather only the blackbody fits to the \swift epochs re-analyzed here.

Some of the objects used for the relationship of \citet{hinkle20a} did not have \swift data near peak and thus have not been re-analyzed in this paper. However, for each of the objects that have been re-analyzed in this work, we followed the procedure of \citet{hinkle20a} to place them on the peak-luminosity/decline-rate relationship plot. In brief, this includes measuring the peak luminosity and time of peak as well as the decline in log luminosity between the peak and 40 days after peak. Using the updated bolometric light curves, we again find that 40 days minimizes the intrinsic scatter in the relationship. For complete details on the analysis and uncertainty computations see \citet{hinkle20a}. We have also updated the classifications of the sources Gaia19bpt, iPTF16fnl, ZTF19aapreis, and iPTF15af based on the improved bolometric corrections. Following the procedure of \citet{kelly07}, we fit this relationship with a linear function, and obtain a best fit of

\begin{equation}
\begin{split}
\log_{10}(L_{peak} / (\text{erg s}^{-1})) = (44.0^{+0.1}_{-0.1}) + \\ (1.5^{+0.3}_{-0.2})(\Delta L_{40} + 0.5)
\end{split}
\end{equation}

\noindent which is consistent with the result of \citet{hinkle20a}. The estimated intrinsic scatter in the peak luminosities of $0.19^{+0.17}_{-0.13}$ dex is significantly decreased from the scatter of $0.29^{+0.23}_{-0.17}$ dex in \citet{hinkle20a}. To examine the statistical significance of this correlation, we performed the Kendall tau test and found a moderately strong correlation of $\tau = 0.46$ with a p-value of $3.3 \times 10^{-3}$. We therefore recover the correlation of \citet{hinkle20a} with decreased scatter and higher significance. We attribute the reduced scatter to the corrected UV photometry, uniform host subtraction procedures, and consistent distances.

\begin{figure*}
\centering
 \includegraphics[width=1.0\textwidth]{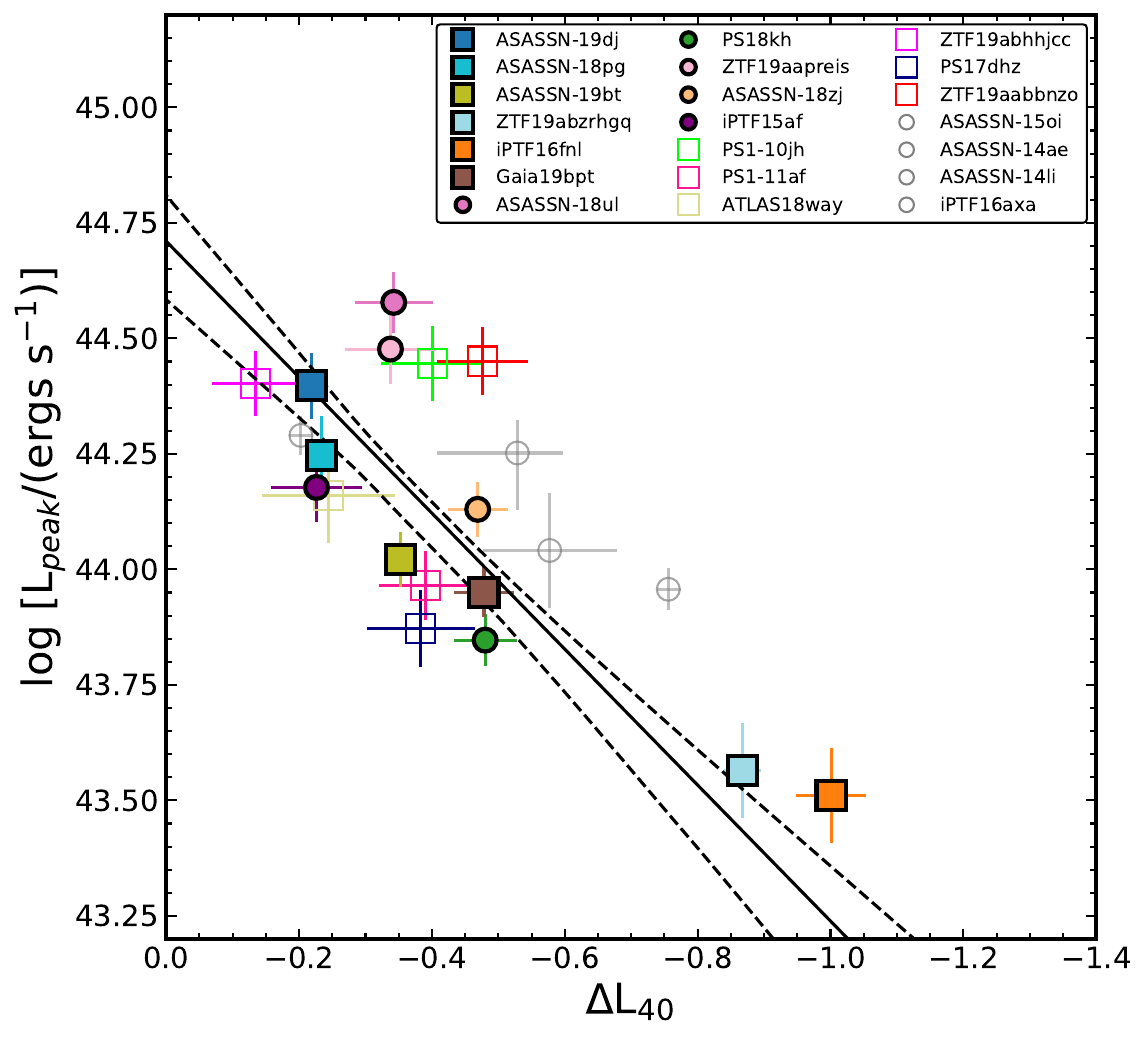}\hfill
 \caption{Peak bolometric UV/optical luminosity as compared to the decline rate $\Delta L_{40} = $ log$_{10}(L_{40}/ L_{peak})$, where $L_{40}$ is the luminosity of the TDE at 40 days after peak. The colors are the same as \citet{hinkle20a} for ease of direct comparison. Following \citet{hinkle20a}, the filled squares with a black border are the ``A'' sample, filled circles with a black border are the ``B'' TDEs, open squares are the ``C'' TDEs, and the gray open circles are the ``D'' TDEs. The solid black line is the line of best fit and the dashed black lines represent plus/minus $1\sigma$ from the best-fit line. The Class ``D'' objects were not included in the fit.}
 \label{fig:L40}
\end{figure*}

\section{Summary} \label{sec:summary}
Following the November 2020 announcement of an updated UVOT calibration to correct for the loss of sensitivity over time, we re-analyzed the published photometry for 37 nuclear outbursts and the ambiguous source ATLAS18qqn. Starting from UVOT images, we re-computed \swift photometry, uniformly modeled the host galaxy SEDs with \textsc{FAST}, corrected the transient photometry for host flux and Galactic extinction, and fit the data with blackbody models. We provide tables of the raw and corrected \swift photometry of the transient, the observed and modeled host-galaxy photometry, the host-galaxy model parameters, and the blackbody models of the transients.

With our updated bolometric UV/optical light curves, we verify the relationship found by \citet{hinkle20a}, that more luminous TDEs decay more slowly than less luminous TDEs. With our uniform data analysis, the scatter in the relationship is significantly reduced.

Given the increased detection rate of TDEs and other exotic transients in recent years, the UV remains a vital wavelength range for studying the transient universe. In particular, UV photometry is a powerful tool for probing the regions close to SMBHs as sources evolve. As more and more similar events are found, \swift will continue to be a key tool in understanding their high-energy emission.

\acknowledgments
We thank the referee for helpful comments that have improved this paper. We also thank Christopher Kochanek for helpful discussions. J.T.H. and this work was supported by NASA award 80NSSC21K0136. Support for T.W.-S.H. was provided by NASA through the NASA Hubble Fellowship grant HST-HF2-51458.001-A awarded by the Space Telescope Science Institute, which is operated by the Association of Universities for Research in Astronomy, Inc., for NASA, under contract NAS5-26555.  B.J.S is supported by NSF grants AST-1908952, AST-1920392, AST-1911074, and NASA award 80NSSC19K1717.

Parts of this research were supported by the Australian Research Council Centre of Excellence for All Sky Astrophysics in 3 Dimensions (ASTRO 3D), through project number CE170100013.

This research has made use of the SVO Filter Profile Service (http://svo2.cab.inta-csic.es/theory/fps/) supported from the Spanish MINECO through grant AYA2017-84089.

\facilities{\swift(UVOT) \citep{roming05}}
\software{linmix \citep{kelly07}}

\bibliography{bibliography}{}
\bibliographystyle{aasjournal}

\end{document}